\documentclass[showpacs,aps,amsmath,amssymb]{revtex4}
\usepackage{graphicx}
\usepackage{dcolumn}
\usepackage{bm}
\usepackage{graphicx}
\usepackage{amsmath}
\begin{document}
\title{Holonomy Transformation in the FRW Metric}
\author{~A.~M. de~M. Carvalho$^{1}$ and Claudio Furtado$^{2}$}
\affiliation{
$^{1}$Departamento de F\'{\i}sica, Universidade Estadual de Feira de Santana, BR116-Norte, Km 3, 44031-460, Feira de Santana, BA, Brazil\\
$^{2}$Departamento de F\'{\i}sica, CCEN,  Universidade Federal da Para\'{\i}ba, Cidade Universit\'{a}ria, 58051-970 Jo\~ao Pessoa, PB, Brazil\\}

\begin{abstract}
In this work we investigate loop variables in Friedman-Robertson-Walker spacetime. We analyze the parallel transport of vectors and spinors in several paths in this spacetime in order to classify its global properties. The band holonomy invariance is analysed in this background.

\end{abstract}


\maketitle
\section{Introduction}
In the loop space formalism for gauge theories the fields depend more on the paths rather than on spacetime points. The fundamental quantity that arises from this path-dependent approach is the non-integrable phase factor that represents a gauge field more adequately than the field strength does. In this approach, electromagnetism, for example, is
a gauge-invariant manifestation of the non-integrable phase factor which gives the exact description of the theory, differently from the field strength and the integral of the potential. As another example of the path-dependent formalism we can cite some aspects concerning phases of gauge fields and confinement of quarks~\cite{pr:wilson}. The extension of the loop space formalism to the theory of gravity was firstly considered by Mandelstam~\cite{Mandelstam} who established several equations involving loop variables, and also by Yang~\cite{Yang}, Voronov and Makeenko ~\cite{Voronov}.

Mathematically speaking, holonomy is a matrix that represents the parallel transport of vectors, spinors, tensors, etc. This matrix provides information on the curvature and topology of a given manifold.  The holonomy matrix can be written as
\begin{equation}
\label{phase}
U_{AB}(C)={\cal P}\left(
-\int_{A}^{B} \Gamma_{\mu}(x(\lambda))\frac{dx^{\mu}}{d\lambda}d\lambda
\right),
\end{equation}
where $\Gamma_{\mu}$ is the tetradic connection and $A$ and $B$ are the initial and final points of the path. Then, associated with every path $C$, from a point $A$ to a point $B$, we have a loop variable $ U_{AB}$ given by ~(\ref{phase}) which, by construction, is a function of the path $C$ as a geometrical object.

Using holonomy, Bollini, Giambiagi and Tiomno~\cite{lanc:bgt} investigated the Kerr black hole space-time and obtained many properties of this geometry. In a recent article, Rothman, Ellis and Murugan ~\cite{rothman:cqg:2001} investigated holonomy in the Schwarzschild-Droste geometry obtaining very interesting results for a class orbits in this space-time. They found that holonomy in this space-time has a ``quantization'' property denominated band
invariance holonomy.  Other investigations on loop variables in the context of gravitational fields includes the computation of this mathematical object in the Schwarzschild-Droste geometry~\cite{rothman:cqg:2001}, in Taub-NUT
space-time\cite{Bini} and in the background spacetime generated by a rotating black string\cite{Carvalho}. In a recent article the clock effects were investigated by Bini, Jantzen and Mashhoon~\cite{cqg:bjm}, via holonomy transformation.
Burges~\cite{prd:b} and Bezerra~\cite{prd:v} examined the effects of a parallel transport of vectors and spinors both around a point-like solution and a cylindrically symmetric cosmic string. This procedure  gives, in general, non-trivial results. These effects point out to the gravitational analogue Aharonov-Bohm effect. One of us (CF)~\cite{gc:afbm} investigated by holonomy transformation, the topological properties of a class of solutions in Kaluza-Klein theory and demonstrated that the holonomy gives a combined effect of gravitation and electromagnetic fields in the parallel transport of vectors in that spacetime. This combined the electromagnetic and gravitational Aharonov-Bohm effects. The analysis of Berry's quantum phase in gravitation~\cite{prd:claudassis} was studied in a recent article. The use of holonomy for quantum computation was analyzed in a geometric approach by Pachos and Zanardi~\cite{ijmp:pz}.

Our study of the global properties of the FRW spacetime is done by computing the   orthonormal frame matrix for the parallel transport of a vector along some specific paths. When this path is closed one obtains the holonomy matrix.  When a vector is parallelly propagated along a loop in a manifold $\textsl{M}$, the curvature of the manifold causes the vector, initially at $p\in \textsl{M}$, to appear rotated with respect to its initial orientation in tangent space $T_{p}\textsl{M}$, when it returns to $p$. The holonomy is the path dependent linear transformation $T_{p}\textsl{M}{\rightarrow}T_{p}\textsl{M}$ responsible for this rotation. Positive and negative curvature manifolds, respectively,  yield deficit or excess angles between initial and final vector orientation under parallel transport around such loops. This global property of the manifold can be used as a means of global classification of spacetimes, as pointed out by Rothman {\it et al.}~\cite{rothman:cqg:2001}, who investigated the holonomy in the  Schwarzschild-Droste geometry. The aim of present article is to investigate the loop variables in FRW spacetime with the objective of studying the global properties of this spacetime. We calculate the loop variables associated  to several paths in this background. The same analyzis is considered in the case of the parallel transport of spinor in this spacetime.

\section{Holonomy Transformation in the FRW Geometry}
\label{sec2}
In this section we present the geometric tools necessary to the development of this article. It is well known that the Friedman-Robertson-Walker metric corresponds to the simplest model of the universe. In this model the universe is considered isotropic and homogeneous, Moreover, the mass is uniformly distributed. The line element of the FRW metric is given by
\begin{equation}
\label{fwr}
 ds^{2}=dt^{2} -\frac{R(t)^{2}}{1-kr^{2}}dr^{2}-R(t)^{2}r^{2}
(d\theta^{2}+\sin^{2}\theta d\phi^{2}),
\end{equation}
where $R(t)$ is the scale factor of the universe, and $k$ is the sign of the curvature. The Gaussian curvature is given in terms of the scale factor by the relation
\begin{equation}
K(t)=\frac{k}{R(t)^{2}}.
\end{equation}
where $k = -1, 0, 1$ means negative, zero, or positive curvature, respectively. In other words, when the curvature is negative, we say that the universe is open; when the curvature is null we say that the universe is flat and finally, when the curvature is negative, the universe is closed. If we  admit that the universe is flat, i. e., $k=0$, the metric of FRW (\ref{fwr}) is reduced to
\begin{equation}
\label{fwr0}
ds^{2}=-dt^{2}+R(t)^{2}g_{ij}dx^{i}dx^{j},
\end{equation}
where $g_{ij}$ is the metric of three-dimensional flat spacetime. 

The line element (\ref{fwr}) is described by a dual 1-form basis (co-frame), defined in terms of local tetrad fields by $e^{a}=e^{a}_{\mu}dx^{\mu}$, where
\begin{subequations}
\label{fwr-basis}
\begin{eqnarray}
        e^{0}&=& dt \mbox{,}       \\
        e^{1}&=& \frac{R(t)}{\sqrt{1-kr^{2}}} dr    \mbox{,}  \\
        e^{2}&=& R(t)rd\theta  \mbox{,}   \\
        e^{3}&=& R(t)r \sin\theta d\phi  \mbox{.}
\end{eqnarray}
\end{subequations}
The tetrad field for the 1-form basis (\ref{fwr-basis}) which corresponds to the FRW metric is given by
\begin{eqnarray}
E^{\mu}_{a}=\mbox{diag}\left(1,\frac{\sqrt{1-kr^{2}}}{R(t)},\frac{1}{R(t) r},\frac{1}{R(t)r\sin \theta}\right).
\end{eqnarray}
We can introduce an affine spin connection 1-form $\omega^{a}_{\mu}$ and define the torsion 2-form and the curvature 2-form, respectively, by
\begin{subequations}
\begin{eqnarray}
T_{a}&=&\frac{1}{2}T^{a}_{bc}\;e^{b}\wedge e^{c}=de^{a}+\omega^{a}_{b} \wedge e^{b}\mbox{,} \\
R^{a}_{b}&=&\frac{1}{2}R^{a}_{bcd}\; e^{c}\wedge e^{d}.
\end{eqnarray}
\end{subequations}
These equations are called Maurer-Cartan structure equations. Using the torsion-free condition for this spacetime, the first of the Maurer-Cartan equations becomes
\begin{equation}
de^{a}+\omega^{a}_{\mu} \wedge e^{b}=0.
\end{equation}
So we can determine the non null 1-form connections. From the 1-form basis (\ref{fwr-basis}) we find the following connections:
\begin{subequations}
\begin{eqnarray}
\omega^{0}_{1}&=&\omega^{1}_{0}=H(t)\; e^{1}      \mbox{,} \\
\omega^{1}_{2}&=&-\omega^{2}_{1}=-\frac{\sqrt{1-kr^{2}}}{Rr} e^{2}    \mbox{,} \\
\omega^{0}_{2}&=&\omega^{2}_{0}=H(t)\; e^{2}      \mbox{,} \\
\omega^{1}_{3}&=&-\omega^{3}_{1}=-\frac{\sqrt{1-kr^{2}}}{Rr} e^{3}\mbox{,}\\
\omega^{0}_{3}&=&\omega^{3}_{0}=H(t)\; e^{3}      \mbox{,} \\
\omega^{2}_{3}&=&-\omega^{3}_{2}=-\frac{\cot\theta}{Rr} e^{3}
\mbox{,}
\end{eqnarray}
\end{subequations}
where $H(t)$ is the Hubble parameter, it express the normalized rate of expansion of the universe, which is given by $\dot{R}/R$. The other terms of the connection are null. The connection 1-forms transforms in the same way as the gauge potential of a non-Abelian gauge theory, which means that any two elements of the group does not commute.

The above connections lead to the following spin connections
\begin{eqnarray}
\Gamma_{\phi}=\left(
\begin{array}{cccc}
0 & 0 & 0 & \dot{R}r\sin\theta \\
0 & 0 & 0 & -\sqrt{1-kr^{2}}\sin\theta   \\
0 & 0 & 0 & \cos\theta \\
\dot{R}r\sin\theta & \sqrt{1-kr^{2}}\sin\theta & -\cos\theta & 0
\end{array}
\right),
\end{eqnarray}

\begin{eqnarray}
\Gamma_{\theta}=\left(
\begin{array}{cccc}
0 & 0 & \dot{R}r & 0 \\
0 & 0 & -\sqrt{1-kr^{2}} & 0  \\
\dot{R}r & \sqrt{1-kr^{2}} & 0 & 0  \\
0 & 0 & 0 & 0
\end{array}
\right),
\end{eqnarray}
and 
\begin{eqnarray}
\Gamma_{r}= \left(
\begin{array}{cccc}
0 & \dot{R}/\sqrt{1-kr^{2}} & 0 & 0 \\
\dot{R}/\sqrt{1-kr^{2}} & 0 & 0 & 0  \\
0 & 0 & 0 & 0  \\
0 & 0 & 0 & 0
\end{array}
\right).
\end{eqnarray}
The spin connection $\Gamma_{\phi}$ corresponds to closed curves with the coordinates $\theta$ and $r$ constants, and the spin connection $\Gamma_{\theta}$ corresponds to orbits with the coordinates azimuthal and radial constants, and finally, $\Gamma_{r}$ which corresponds to orbits with $\phi$ and $\theta$ coordinates constants. The holonomy matrix associated with the parallel transport of vectors around closed curves $\gamma$ is defined by
\begin{equation}
U(\gamma)= {\cal P}\exp \left[ -\oint_{\gamma}dx^{\mu} \Gamma_{\mu} \right],
\end{equation}
where ${\cal P}$ is the order operator. The path $\gamma$ will be restricted only to  circular orbits centered at the origin. We will deal with the transport of vectors along geodesic orbits. The set of all holonomy matrices forms the holonomy group. The holonomy group is defined as the group of linear transformations of the tangent space $T_{p}M$ induced by parallel transport around loops based at a point $p$. This group possesses information regarding to the curvature of the spacetime. 

\subsection{Holonomy Transformation in a Static Universe}
Initially, for pedagogical reasons, we will to deal with the case in which the scale factor $R$ is a constant. In this case the spacetime in question is static.  Moreover, the variables $(r,\theta)$ are fixed along the orbit. This assumptions implies that the holonomy matrix possess only the azimuthal contribution of the spin connection and is written as
\begin{equation}
U(\gamma_{\phi})={\cal P}\exp \left(-\oint \Gamma_{\phi}d\phi \right).
\end{equation}
As the FRW metric is independent of the coordinate $\phi$, the spin connection matrix $\Gamma_{\phi}$ is constant along the path. Expanding in Taylor series the exponential part of the holonomy and reorganizing the odd and even powers terms, we can rewrite the integral holonomy as

\begin{eqnarray}
U(\gamma_{\phi})=1-\frac{\Gamma_{\phi}}{\sqrt{1-kr^{2}\sin^{2}\theta}}\sin\left(
2\pi\sqrt{1-kr^{2}\sin^{2}\theta}\;\right)+\frac{\Gamma_{\phi}^{2}}{1-kr^{2}\sin^{2}\theta}.
\left[1-\cos\left(2\pi\sqrt{1-kr^{2}\sin^{2}\theta}\;\right)
\right]
\end{eqnarray}
Considering only orbits in the equatorial plane, we can write the holonomy matrix as
\begin{eqnarray}\label{stat}
U(\gamma_{\phi})=\left(
\begin{array}{cccc}
1 & 0 & 0 & 0 \\
0 & \cos 2\pi\sqrt{1-kr^{2}} & 0 & \sin 2\pi\sqrt{1-kr^{2}}  \\
0 & 0 &1 & 0  \\
0 & -\sin 2\pi\sqrt{1-kr^{2}} & 0 &  \cos 2\pi\sqrt{1-kr^{2}}
\end{array}
\right).
\end{eqnarray}
This matrix can be interpreted as the rotation generator and the term $2\pi\sqrt{1-kr^{2}}$ means the deficit angle, obtained when we compare the final and the initial positions of the parallelly transported vector. If the holonomy matrix is equal to the identity matrix, it implies that the universe is plane. The parallel transport can be used as a tool to determine the bending of the universe, since the effect of the angle of deficit in this case, is provided only by the curvature of the spacetime. 

The deficit angle $\chi$ obtained when we compare the final and initial position of the parallelly transported vector is given by $\cos \chi_{a}= U^{a}_{a}$, where $a$ is a tetradic index. The terms of non-vanishing angular deviations occur when
$a = 1$ and $2$, so we have
\begin{equation}\label{angu}
\cos\chi_{1 or 2} = \cos 2\pi\sqrt{1-kr^{2}},
\end{equation}
or
\begin{equation}\label{angu1}
|\chi_{1 or 2}| = |2\pi\sqrt{1-kr^{2}}+2\pi n|.
\end{equation}
The above expression shows that for $\chi \neq 0$, if we parallelly transport a vector around a closed path, the final vector does not coincide with the original vector. This physical effect can be understood as a gravitational analogue of the Aharonov-Bohm effect. We saw in (\ref{angu}) that there will be no Aharonov-Bohm effect if $2\pi\sqrt{1-kr^{2}}$ is an integer. This conditions is not always satisfied, because $\sqrt{1-kr^{2}}$ is not necessarily an integer. In fact, this term can assume an arbitrary value. There are  similarities of this effect with the Aharonov-Bohm effect in the cosmic string case~\cite{prd:v}. The crucial difference occurs in the fact that the spacetime exterior to 
the cosmic string has Riemann tensor null  contrary to the FRW  for the  case $k=\pm1$  that has  curvature in the exterior region.

Recently, Rothman {\it et al.} \cite{rothman:cqg:2001} studied some particular curves when a vector ${\bf A}$ is parallelly transported in the geometry of Reissner-Nordstr\"om and then observed very interesting properties of parallel transport that they denominated  band holonomy invariance. We can investigate the band holomy invariance, as proposed by Rothman {\it et al.}, for the present path. The band holomy invariance occurs in a special radius (critical radius) where the holonomy is trivial. In this situation, the invariance of holonomy occurs when $2\pi\sqrt{1-kr^{2}\sin^{2}\theta}\;n=2\pi m$, where $n$ is the number of loops and $m$ is an integer. Using this condition we obtain the the following critical radius
\begin{equation}
\bar{r}=\sqrt{\frac{1-(m/n)^{2}}{k\sin^{2}\theta}}.
\end{equation}
As the critical radius is a positive quantity, it implies that when the universe is open the integers $n$ and $m$ should satisfy the inequality $m>n$, and when the universe is closed, $m<n$. This result suggests a kind of ``quantization" of the holonomy, in which the closing of the orbit depends on the radius of the orbit. This feature indicates the existence of curvature in the region of the spacetime enclosed by the orbit. We notice that  there will be no Aharonov-Bohm effect if the radius of the orbit is equal to the critical radius. We can always choose an orbit in which the holonomy is trivial.

\subsection{Non Static Universe}

If we consider the non static universe, the scale factor is not any longer a constant so the universe can be assumed to be in expansion. Let's determine the holonomy matrices for a function $R(t)$. For simplicity, we assume that the circular orbits are in the equatorial plane. The holonomy transformation is evaluated in the same way as in the previous case. The azimuthal contribution for the holonomy is now given by
\begin{eqnarray}\label{expandingh}
U(\gamma_{\phi})=1-\frac{\Gamma_{\phi}}{\sqrt{1-(k+\dot{R}^{2})r^{2}}}\sin\left(2\pi\sqrt{1-(k+\dot{R}^{2})r^{2}}
\;\right)+\frac{\Gamma_{\phi}^{2}}{1-(k+\dot{R}^{2})r^{2}}
\left[1-\cos\left(2\pi\sqrt{1-(k+\dot{R}^{2})r^{2}}\;\right)
\right].
\end{eqnarray}
We can also study the bands of invariance of holonomy. In the above expression, the band of invariance corresponds to the relation $2\pi n\sqrt{1-(k+\dot{R}^{2})r^{2}}=2\pi m$. Henceforth, we find the following value for the critical radius
\begin{equation}\label{critt}
\bar{r}= \sqrt{\frac{1-(m/n)^{2}}{k+\dot{R}^{2}}}.
\end{equation}
Now, the critical radius depends on the form of the derivative of the scale factor $R(t)$. For finite $r$, given by (\ref{critt}), the holonomy is again the identity matrix. This implies that, for some critical radii, after $n$ loops, the transported vector does not acquire a deficit angle. The deficit angle $\chi$ obtained when we compare the final and initial position of the parallel transported vector is given by $\cos \chi_{a}= U^{a}_{a}$, where $a$ is a tetradic index. The terms of non-vanishing angular deviations occur when
$a = 1$ and $2$. The deficit angle $\chi$ in this case is given by
\begin{eqnarray}\label{angu3}
\chi=2\pi \sqrt{1-(k+\dot{R}^{2})r^{2}}.
\end{eqnarray}

Now, the deficit angle depends on the scale factor $R(t)$. In order to check this result, when we consider $\dot{R}=0$ we obtain the same result of the static universe. Note that this behaviour is similar to the static case, if we parallelly transport a vector around a closed path, the final vector does not coincide with the original vector.
This physical effect could be understood as a gravitational analogue, for parallel transport of vectors in a universe in expansion, of the Aharonov-Bohm effect. We can see in  expression (\ref{angu3}), in the case of a flat, $k=0$, but expanding universe the holonomy is nontrivial. For $k=0$ the tri-dimensional spacial section of FRW spacetime is flat, the holonomy matrix (\ref{expandingh}) is the transport parallel matrix for a closed circular constant time curve in the four-dimensional expanding universe. In this way, holonomy experiences the effect of expansion of the universe.  The expression (\ref{angu3}) gives, in principle, a way to measure the expansion rate of the universe, its gives a relation between the deficit  angle $\chi$ and expansion rate $\dot{R}$, the size of this effect can be evaluated, if we consider a realistic  problem  that envolves the parallel transport of vectors.  If we consider a constant time circular orbit in the equatorial plane of the part of FRW spacetime corresponding to three-space,  we obtain the following expression for the  holonomy matrix
\begin{eqnarray}\label{tridiemsional}
U(\gamma_{\phi})=1-\frac{\Gamma_{\phi}}{\sqrt{1-kr^{2}}}\sin\left(
2\pi\sqrt{1-kr^{2}}\;\right)+\frac{\Gamma_{\phi}^{2}}{1-kr^{2}}
\left[1-\cos\left(2\pi\sqrt{1-kr^{2}}\;\right)
\right],
\end{eqnarray}

where $\Gamma_{\phi}$ is the tridimensional spin connection  given by
\begin{eqnarray}
\Gamma_{\phi}=\left(
\begin{array}{ccc}
 0 & 0 & -\sqrt{1-kr^{2}}   \\
 0 & 0 & 0 \\
 \sqrt{1-kr^{2}} & 0 & 0
\end{array}
\right),
\end{eqnarray}
Note that in this case the holonomy matrix is non-trivial for $k=\pm 1$. For $k=0$ the holonomy matrix is trivial, which is consistent with fact of the scalar curvature is zero in this case.

\section{Spinorial Parallel Transport}
In this section we are interested in the study of the parallel transport of a spinor in the FRW geometry. An important question that emerges when we study the parallel transport of vectors is what happens with more complex fields when they also undergo a parallel transport. We know that vectors when parallelly transported in a curved background will be changed after a complete loop, if there is a nonvanishing curvature in the region surrounded by the loop. So, we are interested  in understanding what comes about when spinors are parallelly transported in the FRW geometry. The spinorial derivative is given in terms of the spinorial connection, defined as
\begin{equation}
\label{holonomy2}
\Gamma_{\mu}(x)=-\frac{1}{4}\omega^{\alpha}_{\nu \mu}\gamma_{\alpha}\gamma^{\nu},
\end{equation}
where $\gamma_{\alpha}$ are flat-space Dirac matrices and $\omega^{\alpha}_{\nu \mu}$ are the coefficients of the spin connection. The holonomy matrix is given in terms of the spinorial connection $\Gamma_{\mu}(x)$ instead of the spin connection, by the following expression
\begin{equation}
U(\gamma)= {\cal P}\exp \left[ -\oint_{\gamma}dx^{\mu} \Gamma_{\mu}(x) \right].
\end{equation}

\subsection{Static Universe}

For circular orbits with constant time we found the following expression for the spin connection in terms of the Paulis spin matrices
\begin{eqnarray}
\Gamma_{\phi}(x)&=&\frac{1}{4}\sqrt{1-kr^{2}}\sin\theta\left(\gamma_{1}\gamma^{3}-\gamma_{3}\gamma^{1}
\right)\nonumber \\
&=&\frac{1}{2} \sqrt{1-kr^{2}}\sin\theta
\left(
\begin{array}{cc}
0 & 1 \\
-1& 0
\end{array}.
\right).
\end{eqnarray}
The phase associated with the above spinor connection is given by
\begin{eqnarray}
U(\gamma_{\phi})&=&\exp\left(-\oint \Gamma_{\phi}(x)d\phi \right)\nonumber \\
        &=& \exp \left(-2\pi \Gamma_{\phi}(x)\right)\mbox{.}
\end{eqnarray}
Expanding this term we obtain
\begin{equation}
U(\gamma_{\phi})=1-\frac{\Gamma_{\phi}(x)}{\sqrt{1-kr^{2}}\sin\theta/2}\sin(\pi
\sqrt{1-kr^{2}}\sin\theta)-
\frac{\Gamma_{\phi}(x)^{2}}{(\sqrt{1-kr^{2}}\sin\theta/2)^{2}}\left[1-\cos(\pi\sqrt{1-kr^{2}}\sin\theta
)\right].
\end{equation}
Or, in a matricial form
\begin{eqnarray}
U(\gamma_{\phi})=\left(
\begin{array}{cc}
\cos(\pi\sqrt{1-kr^{2}}\sin\theta) & -\sin(\pi\sqrt{1-kr^{2}}\sin\theta) \\
\sin(\pi\sqrt{1-kr^{2}}\sin\theta) &
\cos(\pi\sqrt{1-kr^{2}}\sin\theta)
\end{array}
\right).
\end{eqnarray}
This matrix corresponds to the phase acquired by a spinor when it is parallelly transported along a closed orbit with a fixed radius. Note that for $k=0$ the holonomy is trivial.

For orbits with constant azimuthal angle, the unique non-null term is
\begin{eqnarray}
\Gamma_{\theta}(x)&=&-\frac{1}{4}\sqrt{1-kr^{2}}\left(\gamma_{0}\gamma^{1}+\gamma_{1}\gamma^{0} \right)\nonumber \\
&=&\frac{1}{2}\sqrt{1-kr^{2}}\; i\sigma_{3}.
\end{eqnarray}
The expression for the holonomy is obtained similarly to the previous cases, so the $\Gamma(x)_{\theta}$ contribution to the holonomy matrix is given by
\begin{eqnarray}
\label{spintemp}
U(\gamma_{\theta})=1-\frac{\Gamma_{\theta}}{\sqrt{1-kr^{2}}/2}\sinh\left(\pi
\frac{\sqrt{1-kr^{2}}}{2}\right) +
\frac{(\Gamma_{\theta})^{2}}{(1-kr^{2})/4}\left[\cosh\left(\pi
\frac{\sqrt{1-kr^{2}}}{2}\right) -1 \right].
\end{eqnarray}

We return to the analysis of constant time circular orbits. Instead of vectors, we will study the parallel transport of spinors. In order to do this, we will make the covariant spinorial derivative null
\begin{equation}
\partial_{\mu}\psi -\Gamma_{\mu}(x)\psi=0.
\end{equation}
The first case to be analyzed will be that of orbits with constant $\theta$. In this case the unique non-null spinorial connection is $\Gamma_{\phi}$, thus, we will have the following system to solve for the two-component spinor
\begin{subequations}
\begin{eqnarray}
\frac{\partial \psi_{0}}{\partial \phi}-\frac{1}{2}\sqrt{1-kr^{2}}\sin\theta  \psi_{1}&=&0, \\
\frac{\partial \psi_{1}}{\partial
\phi}+\frac{1}{2}\sqrt{1-kr^{2}}\sin\theta  \psi_{0}&=&0 .
\end{eqnarray}
\end{subequations}
These equations are easily integrated to
\begin{subequations}
\begin{eqnarray}
\psi_{0}&=&\psi_{0}(0)\cos \omega_{\phi}\phi - \psi_{1}(0)\sin \omega_{\phi}\phi, \\
\psi_{1}&=&\psi_{0}(0)\sin \omega_{\phi}\phi + \psi_{1}(0)\cos
\omega_{\phi}\phi,
\end{eqnarray}
\end{subequations}
with $\omega_{\phi}=\sqrt{1-kr^{2}}\sin\theta /2$. This result shows that, when the spinor is parallelly transported, its components are changed from $\psi(0)$ to $\psi$. 

We will analyze another trajectory, which corresponds to orbits with the azimuthal angle fixed. Thus, we need to solve the following system of equations
\begin{subequations}
\begin{eqnarray}
\frac{\partial\psi_{0}}{\partial\theta}-\omega_{\theta}i  \psi_{0}&=&0, \\
\frac{\partial\psi_{1}}{\partial\theta}+\omega_{\theta}i
\psi_{1}&=&0 ,
\end{eqnarray}
\end{subequations}
with $\omega_{\theta}=\sqrt{1-kr^{2}}/2$. The solution is given by
\begin{subequations}
\begin{eqnarray}
\psi_{0}&=&\psi_{0}\exp( i\omega_{\theta}\theta), \\
\psi_{1}&=&\psi_{1}\exp( -i\omega_{\theta}\theta).
\end{eqnarray}
\end{subequations}
The behaviour of the solutions depends on the sign of the ``frequency" $\omega_{\phi}$ or $\omega_{\theta}$. If they are positive, the solutions will be oscillatory. If not, the trigonometric solutions will be replaced by hyperbolic functions.
\subsection{Non-static Universe}
If we consider a non-static universe the azimuthal contribution to the spinorial connection is given by
\begin{eqnarray}
\Gamma_{\phi}(x)&=&\frac{1}{2}\dot{R}r\sin\theta\; \gamma_{0}\gamma_{3} +\frac{1}{2}\sqrt{1-kr^{2}}\sin\theta\;\gamma_{3}\gamma_{1}+\frac{1}{2}\cos\theta\; \gamma_{3}\gamma_{2}.
\end{eqnarray}
If we consider only the equatorial case, the holonomy matrix can be evaluated directly from Eq. (\ref{holonomy2})
\begin{equation}
U(\gamma_{\phi})=1-\frac{\Gamma_{\phi}(x)}{\sqrt{1-(\dot{R}^{2}+k)r^{2}}}\sin(\pi
\sqrt{1-(\dot{R}^{2}+k)r^{2}})-
\frac{(\Gamma_{\phi}(x))^{2}}{1-(\dot{R}^{2}+k)r^{2}}\left[1-\cos(\pi\sqrt{(1-(\dot{R}^{2}+k)r^{2}})\right].
\end{equation}
Where we write the spinorial connection as
\begin{eqnarray}
\Gamma_{\phi}(x)=\frac{1}{2}
\left(
\begin{array}{cc}
A & B\\
-B & -A
\end{array}
\right),
\end{eqnarray}
where $A=\dot{R}r/2$ and $B=\sqrt{1-kr^{2}}/2$. When a spinor under goes a parallel transport around a closed curve, if the final result differs of the original spinor, it is due to global effects of the curvature in the interior region. 
The critical radius depends not only on the curvature, it also depends on the derivative of the scale factor $\dot{R}(t)$
\begin{equation}
\bar{r}= \sqrt{\frac{1-(2m/n)^{2}}{k+\dot{R}^{2}}}.
\end{equation}

\section{Concluding Remarks}
The holonomy matrix gives the parallel transport of a vector along a given path. It is frequently used as a means of studying global characteristics of spacetimes. In this article we use it to investigate the Friedman-Robertson-Walker spacetime. When analysing constant time circular orbits we find some critical radii, where the holonomy matrix is trivial, implying that the parallel transport of a vector along such orbit does not change its orientation. Comparing with  the Aharonov-Bohm (AB) phase $\oint A_{\mu}dx^{\mu}$, 
the gravitational analogue AB effect~\cite{prd:v} occurs when the holonomy  $\oint \Gamma_{\mu}dx^{\mu}$  is 
nontrivial. Therefore,  there is a analogue AB effect in Friedman-Walker-Robertson spacetime , except at those
 critical radii. The holonomy is not trivial in the static case, except for those critical radii and for the $k=0$ case.
 In the non-static case the holonomy is not trivial, except for critical radii where the band holonomy is observed. In the expanding universe the holonomy depends on the derivative of the scale factor. We have shown, by explicit computation from the metric corresponding to a Friedman-Robertson-Walker universe, that the loop variables are a combination of 
rotations and boosts in this spacetime.  We have demonstrated that the holomy is trivial in this spacetime for
 some special values of the coordinate parameter\cite{rothman:cqg:2001} that implies in holonomy quantization. Similar  effects were investigated  in a series of black-hole solutions\cite{rothman:cqg:2001,Carvalho,jhep:c,ijmpd:c}. The holonomy matrix for spinors was analyzed in constant time circular  orbits in Friedman-Walker-Robertson spacetime. We have found that the holonomy is not trivial in the static case, except for $k=0$ and for critical radii where a band holonomy is observed. In a non-static universe the holonomy depends on the expansion parameter and is non-trivial, except for the case  mentioned in the non-static case of a parallel transport of a vector. Finally, we remark the importance of the study of  the holonomy matrix, which  can be used to investigate, in the geometric approximation, the parallel transport of  the polarization vector of light. In this approximation, in principle, one could measure polarization bands in the present spacetime.

{\bf Aknowledgement} \quad We thank CAPES (PROCAD), CNPq/Pronex/FAPESQ-PB and CNPQ/Projeto Universal  for
financial support.



\begin{thebibliography}{99}

\bibitem{pr:wilson}
K.  Wilson,  Phys. Rev. D {\bf 10}, 2445 (1964).

\bibitem{Mandelstam}
S. Mandelstam,  Ann. Phys. {\bf 19}, 25 (1962);
Phys. Rev. {\bf 175}, 1604 (1968).

\bibitem{Yang} C. N. Yang, Phys. Rev. Lett. {\bf 33}, 445 (1974) .

\bibitem{Voronov}
N. A. Voronov   and  Y. M. Makeenko,  Sov. J. Nucl. Phys., {\bf 36}, 444 (1982).

\bibitem{lanc:bgt}
C. G. Bollini, J. J. Giambiagi and J. Tiomno.  Lett. Al Nuovo Cimento, 31, (1) (1981).

\bibitem{rothman:cqg:2001}
Tony Rothman, George Ellis and Jeff Murugan. Class. Quant. Grav. {\bf 18}, 1217 (2001).

\bibitem{Bini}
D. Bini,  C. Cherubini and R. T. Jantzen, Class. Quantum Grav. {\bf 19}, 5481 (2002).

\bibitem{Carvalho}
A. M. de M Carvalho ,  F. Moraes, and  C. Furtado,  Class. Quantum Grav. {\bf 20}, 2063 (2003).

\bibitem{cqg:bjm}
D. Bini,  R. Jantzen and B. Mashhoon, Class. Quant. Grav. {\bf 19} 17 (2002).

\bibitem{prd:b}
C. J. C. Burges, Phys. Rev. D, {\bf 32}, 02 (1985).

\bibitem{prd:v}
V. B. Bezerra,  Phys. Rev. D,  {\bf 35}, 06 (1987).

\bibitem{gc:afbm}
J. G. Assis, C. Furtado, V. Bezerra and F. Moraes, Gravitation $\&$ Cosmology 6, 233 (2000).

\bibitem{prd:claudassis}
J. G. Assis, C. Furtado and V. B. Bezerra, Phys. Rev. D {\bf 62}, 45003 (2000).

\bibitem{ijmp:pz}
J. Pachos  and  P. Zanardi, Int.J.Mod.Phys. {\bf B15}, 1257 (2001).

\bibitem{atmp:mal}
J. Maldacena, Adv. Ther. Math. Phys. {\bf 2}, 231 (1998).

\bibitem{atmp:wit}E. Witten, Adv. Ther. Math. Phys. {\bf 2} 253 (1998).

\bibitem{jhep:c}A. M. de M. Carvalho, F. Moraes and C. Furtado, Mod. Phys. Lett.{\bf A 19}, 2683 (2004). 

\bibitem{ijmpd:c}A. M. de M. Carvalho, F. Moraes and C. Furtado, JHEP 0406, 029 (2004).

\end{thebibliography}
\end{document}